# Electron Transport Properties of Graphene-Graphyne-Graphene Transistors: First Principles Study

*Young I. Jhon*[*][†] *and Myung S. Jhon*[†][‡]

[†]Nano-convergence Core Technology for Human Interface (WCU), School of Advanced Materials Science and Engineering, Sungkyunkwan University, Suwon 440-746, Korea
[‡]Department of Chemical Engineering, Carnegie Mellon University, Pittsburgh PA 15213, USA

## Abstract

A novel nanoelectronic device is constructed by graphyne that is robustly connected between graphene electrodes, where graphyne is composed of hexagonal carbon rings and carbon chains. Owing to similarities between the bond lengths and unit cell shapes of graphene and graphyne, they have perfectly matched interfacial structure at periodic locations, enabling the facilitated charge transfer and heterostructural stability. Using a combined nonequilibrium Green's function and density functional theory formalism, we have systematically investigated the electron transport properties of graphene-graphyne-graphene field effect transistors (FETs) by varying the graphyne size and the carbon chain length. These devices exhibit excellent switching behaviors with ON/OFF ratios on the order of $10^2$–$10^3$. The ON/OFF ratio increases as either of the graphyne size or the carbon chain length increases. Noticeably, these devices sustain good FET features even at the small graphyne size of 8.5 Å, yielding ON/OFF ratio of 650 and transmission energy gap of 0.8 eV, which suggests their potential applications for fabricating highly-integrated circuits at the level comparable to molecular devices. Their boron-nitrides analogues show similar qualitative behaviors for the changes of the graphyne size and bias voltage, but they show higher ON/OFF ratios for the smaller chain length in contrast to graphyne TFTs.

[*] *Corresponding author.* Fax: +82-31-290-7410, +412-268-7139.
Email address: yijhon@hotmail.com (Y.I. Jhon)

# INTRODUCTION

Graphene, a single atomic carbon film with hexagonal sp$^2$ covalently-bonded structure, has drawn great attention for electronic applications due to its ultrafast electron mobility.[1–4] This superb electronic property results from its massless Dirac fermions. Dirac upper and lower cones consisting of the conduction and the valence bands meet at a point on the Fermi level, thus making graphene a semimetal with zero band gap.[5–7] Among various potential fields of graphene electronics, much effort has been made to developing graphene field effect transistors (FETs) because they are fundamental units of logic integrated circuits. The preceding studies show that graphene FETs can have extremely high carrier mobility in excess of 100,000 cm$^2$ V$^{-1}$ s$^{-1}$, but their exceedingly small ON/OFF ratios, which basically stem from the zero band gap of graphene, have considerably disturbed practical applications of graphene FFTs. To make a sizable band gap in graphene, several methods have been suggested including chemical decorations of graphene surfaces and edges,[8–10] application of an electric field perpendicular to bilayer graphene,[11–13] substrate induced symmetry breaking in graphene,[14] and fabrications of graphene nanoribbons.[15–17]

On the other hand, one of the most serious problems encountered in the fabrication of electronic devices using a monoatomic layer semiconductor is the contact between the source/drain metal and the monoatomic layer channel. For instance, the contact resistance between graphene and metal source/drain electrodes composed of Au(25 nm)/Ti(10 nm) is as high as 450–800 Ω μm, which severely degrades the performance of the device.[18] To solve this problem, researchers began to consider the use of graphene as electrodes that can be

robustly connected with the monoatomic layer semiconductor at the interfacial edges.[18–21]

Recently, the research on graphene has been rapidly expanded to different type of two-dimensional carbon allotrope, *i.e.*, graphyne.[22–25] It is revealed that graphyne can have either of multi-Dirac cones[22] or nonzero band gap[24,25] in the electronic band structure, depending on its specific structure. For instance, $\alpha$-graphynes, which are two-dimensional structures composed of single and triple bonded carbon atoms between the corner atoms of the honeycomb structures, possess electronic properties similar to that of single-layer graphene with respect to the existence of massless Dirac fermions.[22] This indicates the condition that all carbon atoms are chemically equivalent is not a prerequisite for the existence of Dirac point in the electronic structure of graphene. Furthermore, the study on 6,6,12-graphyne suggests that even hexagonal symmetry of graphene is neither a prerequisite for the appearance of Dirac point. Finite-size building blocks of these graphyne structures have already been synthesized.[26–31] Özçelik *et al.* demonstrate that $\alpha$-graphynes can be stable for even $n$ while they are unstable for odd $n$, where $n$ is the number of constituent carbon atoms of the chain.[32]

In contrast to $\alpha$-graphynes that are composed of carbon chains only, quantum mechanical studies indicate that graphynes that are composed of hexagonal carbon rings joined by carbon chains (Figure 1, these graphynes are denoted as *R*-graphynes and unless otherwise stated, graphynes mean *R*-graphenes in this study) should have nonzero band gaps, and the sizes of which can be modulated by varying the carbon ring size or carbon chain length.[24,25] Very recently, graphdiyne, which is *R*-graphene with a chain composed of four carbon atoms (Figure 1b), has been successfully synthesized via cross-linking reaction on top of copper

surface, making an initial step towards the actual use of $R$-graphyne.[33]

In this paper, considering the exceptional electronic mobility of graphene and nonzero band gap of $R$-graphyne along with structural/compositional similarities between graphene and $R$-graphyne, a new kind of nanoelectronic TFTs are proposed where $R$-graphynes is robustly connected between graphene electrodes. The structures of $R$-graphyne and graphene are perfectly matched at periodic locations when they are connected in parallel to the zigzag direction of graphene (Figure 2). Such robust interfacial connections enable the facilitated charge transfer between $R$-graphyne and graphene electrodes as well as providing structural stabilities of their heterojunctions. Using density functional theory calculations combined with nonequilibrium Green's function, we have systematically studied the electron transport properties of these graphene-graphyne-graphene TFTs. Special attention is paid on the carbon chain length effect of $R$-graphyne and the size effect of $R$-graphyne region on the electron transport behaviors of these TFTs.

Throughout this paper, $R$-graphyne with a chain composed of $n$ carbon atoms is denoted as R-C$n$-graphyne (Figure 1) while its device with graphene electrodes is denoted as GRP-$m$GRPY$n$-GRP (Figure 2), where GRPE and GRPY are graphene and graphyne, respectively, $m$ is the number of GRPY units in source-to-drain direction, and $n$ is the number of carbon atoms consisting of a carbon chain ($n$ is 2 or 4 in this study). It is possible to form hexagonal boron-nitride (h-BN) analogues[34−36] of the above graphyne systems and we have examined electronic characteristics of these h-BN analogues as well. The h-BN analogues of R-C$n$-graphyne and GRP-$m$GRPY$n$-GRP are denotes as R-C$n$-BNyne and GRP-$m$BNY$n$-GRP, respectively.

## COMPUTATIONAL METHODS

All quantum mechanical calculations are performed using the ATK ab-initio program.[37,38] In these calculations, Double-zeta single polarized basis sets with the mesh cutoff of 150 Rydberg are used and the generalized gradient approximations (GGA) within Perdue-Burke-Ernzerhof functional[39] are employed along with norm-conserving pseudo potentials for core electrons. In reciprocal space, Brillouin zone is sampled using the Monkhorst-Pack scheme[40] with the k-meshes of $17(13) \times 1 \times 100$ for the systems of $n = 2(4)$ where 100 are used for the integration in the electrodes. The system dimensions are determined based on the optimized graphyne structures.

The electronic currents of FETs are calculated using Landauer-Büttiker equation as given by

$$I = \frac{2e}{h} \int_{-\infty}^{+\infty} T(E, V) \left[ f(E - \mu_{\mathrm{L}}) - f(E - \mu_{\mathrm{R}}) \right] \tag{1}$$

where $e$ denotes the electron charge, $h$ is Planck's constant, $f$ is the Fermi function of electrodes, $\mu_{\mathrm{L}}$ and $\mu_{\mathrm{R}}$ are the chemical potentials of left and right electrodes, respectively, and $T(E, V)$ is transmission probability as a function of electron energy.

## RESULTS AND DISCUSSION

**Electronic Structures.** After obtaining the optimized structures of *R*-C2-graphyne/R-C4/graphene (Figure 1a,b), their electronic band structures are calculated using density functional theory method (Figure 3a,b). The direct band gap emerges at *M*-point for R-C2-graphyne while it emerges at *Γ*-point for R-C4-graphyne and their sizes are evaluated to be 0.65 and 0.70 eV, respectively. The electronic structures of their h-BN-analogues (Figure 1c,d) show that the dispersion curves of h-BN-analogues are much flattened but the direct band gaps are developed at the same k-points compared to the corresponding graphynes (Figure 3c,d). The band gap sizes of R-C2-BNyne and R-C4-BNyne are 2.94 and 1.95 eV, respectively. It is noticeable that the band gap size of graphyne is rather insensitive to the change of the carbon chain length while the band gap size of h-BN-analogue remarkably decreases as the chain length increases, which is in good agreement with the result of Zhou *et al.*'s work although our h-BN-analogues slightly differ from those in their work with respect to chemical compositions and our band gap values are somewhat greater than those obtained from their calculations.

**Graphene-Graphyne-Graphene TFTs.** The electron transport properties of GRP-*m*GRPY2-GRP are investigated where *m* ranges from 1 to 3 in order to examine the size effect of R-C2-graphyne on its FET performance. There exist distinct energy gaps around Fermi level in the electron transmission spectra and the energy gap is evaluated to be ~0.1 eV for *m* = 1 while it is ~0.8 eV for *m* = 2 or 3, suggesting good switching behaviors of these devices (Figure 4a).

On the assumption of the linear response of transmission conductance to the gate potential, in which the effect of the gate potential is a simple shift of the relative positions of the electrode Fermi levels within the transmission spectrum, current-voltage ($I$–$V$) curves of these devices are calculated for the bias voltage of 0.05, 0.10, and 0.15 eV. The $I$–$V$ curve shifts in the bottom right direction as the bias voltage decreases, regardless of the size of R-C2-graphyne region (Figure 5a–c). We attribute this downward shift of the $I$–$V$ curve to the lowered leakage current. In addition, we infer that the left shift of the $I$–$V$ curve as increasing bias voltage should result from the variation in work function difference between R-C2-graphyne and graphene. This left shift of $I$–$V$ curve indicates that the device becomes more heavily n-doped as the bias voltage increases. Under the bias voltage of 0.05 V, ON/OFF ratio of 93 is achieved for GRP-1GRPY2-GRP device and its magnitude increases further to 127 and 113 for GRP-2GRPY2-GRP and GRP-3GRPY2-GRP devices, respectively (Figure 5d).

Additionally, in order to investigate the chain length effect of graphene-graphyne-graphene TFTs, the electron transport properties of GRP-$m$GRPY4-GRP devices are examined where $m$ ranges from 1 to 3. More attention should be paid to these R-C4-grphyne TFTs because R-C4-graphyne has been successfully synthesized experimentally. The transmission energy gaps of these devices are measured to be 0.6–0.8 eV (Figure 4b), and $I$–$V$ curves of these devices exhibit pronounced n-type TFT characteristics (Figure 6a–c). The $I$–$V$ curve shifts in the bottom-right direction as the bias voltage decreases, similarly to GRP-GRPY2-GRP devices. On/OFF ratios of these devices sensitively depend on the size of R-C4-graphyne. The ON/OFF ratio is ~650 for $m = 1$ and this value greatly increases to ~5,100 and ~8,800 for $m = 2$ and 3, respectively (Figure 6d). It is notable that ON/OFF ratio of as high as 650 is achieved for the small graphyne size of 8.5 Å. This indicates that using R-C4-graphyne, we can fabricate decent TFT circuits at an integration level comparable to molecular TFT devices.

**Graphene-BNyne-Graphene Devices.** It is known that hexagonal boron nitride analogues can exist for many structures of graphene family including graphyne. This chemical extension opens another field of monoatomic layer electronics. Most of device characteristics can be diversified by this BN substitution. In this context, we also studied electronic characteristics of TFT devices composed of h-BN analogues of R-C2-graphyne/R-C4-graphyne as illustrated in Figure 7. The energy gap ranges between 2.0 eV and 4.5 eV in the electron transmission spectra (Figure 8a,b), which are decently greater than the energy gap values (0.1–0.8 eV) obtained from corresponding graphyne devices. These values of 2.0–4.5 eV are comparable to the transmission energy gaps (4.0–4.5 eV) shown in graphene-boron nitride-graphene nanoribbon TFTs.[20] It is notable that the transmission energy gaps can be lowered to 2.0–2.5 eV by reducing the sizes of R-C2-BNyne/R-C4-BNyne to the smallest ones of 5.68 and 8.5 Å, respectively.

*I–V* curves of these BN analogues show typical n-doped TFT behaviors, clearly exhibiting monotonic increases of electronic currents as the gate voltage increases (Figure 8c,d). For the bias voltage of 0.01 V, ON/OFF ratio of GRP-1BNY2-GRP device is evaluated to be ~1,100 but it greatly increases to be on the order of $10^4$ and $10^6$ for GRP-2BNY2-GRP and GRP-3BNY2-GRP devices, respectively (Figure 8c). Interestingly, ON/OFF ratios (500-1,400) of R-C4-BNyne TFTs (Figure 8d) are much smaller than those of R-C2-BNyne TFTs although the chain length of R-C4-BNyne is longer than that of R-C2-BNyne. It is opposite to the trend observed in the graphyne TFTs, which may be concerned with the chain length effect on the electronic band gaps of graphyne/BNyne as shown in Figure 3.

Considering overall factors such as system dimensions, TFT performance, and transmission energy gaps etc., we suppose that GRP-1BNY2-GRP would be the most promising candidate for future FET applications among the various graphene-BNyne-graphene devices.

**Electronic Orbital Analyses** To get physical insight into the electronic transport of graphene-graphyne-graphene devices, we investigate HOMO (highest occupied molecular orbital) and LUMO (lowest unoccupied molecular orbital) of the atoms consisting of central regions in these devices. In GRP-1GRPY2-GRP (Figure 9a,b). and GRP-3GRPY2-GRP (Figure 9c,d) devices, these HOMO and LUMO dominantly possess π/π* characteristics of carbon atoms Particularly, it is noted that the spatial distribution of LUMO is driven into inner graphyne region or gathered along source to drain direction, making electronic channels between the two graphene electrodes. On the contrary, the spatial distribution of HOMO is dispersed along the directions other than source to drain direction or gathered in the boundary regions between graphyne and graphene. This pattern is also observed in GRP-3GRPY4-GRP (Figure 9e,f) device, forming π* characteristic channels at the LUMO state.

In contrast that the electrons are remarkably delocalized at HOMO and LUMO in graphyne-based devices, they are localized at one side of the central region in GRP-3BNY2-GRP (Figure 10a,b). Such electronic localizations are shown more at HOMO-1 and LUMO-1 (Figure 10c,d) and at HOMO-2 and LUMO-2 (Figure 10e,f). These results indicate that graphyne TFTs should offer more facilitated electronic mobilities compared to their h-BN analogues.

## 4. CONCLUSIONS

For the first time, graphyne-based TFT devices are suggested where graphyne is connected with graphene electrodes creating perfectly matched interfacial structures at periodic locations. These robust connections enable the facilitated charge transfer between graphyne and graphene electrodes, solving preceding contact problems of metal electrodes.

Using density functional theory calculation combined with nonequilibrium Green's function, we demonstrate that these graphene-graphyne-graphene TFTs can achieve excellent ON/OFF ratio on the order of $10^2$–$10^3$. In particular, it is noted that the graphyne size of these devices can be reduced to below 1 nm while maintaining a good switching behavior, suggesting their potential applications for fabricating integrated circuits at the extremely high integration level comparable to molecular devices. As either of the graphyne size or the carbon chain length increases, the ON/OFF ratio increases in general.

In contrast to graphyne TFTs, their h-BN analogues have augmented energy gaps of 2.0–4.5 eV in the electron transmission spectra and show the opposite chain length effect resulting that ON/OFF ratios of R-C2-BNyne TFTs surpass those of R-C4-BNyne TFTs.

Electronic orbital analysis indicates that electrons in the conduction and valence orbitals are considerably delocalized in graphyne TFTs while the electrons are localized towards the boundaries in their h-BN analogues. This result suggests that graphyne TFTs should offer more facilitated electron mobilities than their h-BN analogues. This study will have important implications to developing graphyne-based monoatomic layer FET devices.

## AUTHOR INOFRMATION

### Corresponding Author

E-mail: yijhon@hotmail.com

## ACKNOWLEDGMENTS

This work was supported by the World Class University program of KOSEF (Grant No. R32-2008-000-10124-0).

**Figure captions**

**Figure 1.** The structures of **(a)** R-C2-graphyne, **(b)** R-C4-graphyene, **(c)** R-C2-BNyne, and **(d)** R-C4-BNyne. Gray, bronze, and blue atoms are carbon, boron, and nitrogen atoms, respectively. The region surrounded by green dotted lines indicates a primitive cell of each system.

**Figure 2.** The structures of **(a)** GRP-$m$GRPY2-GRP and **(b)** GRP-$m$GRPY4-GRP devices ($m$ = 1−3). Grey and white atoms are carbon and hydrogen atoms, respectively. The unit size of $R$-graphyne is denoted by white clamp notation and its number ($m$) ranges over 1−3.

**Figure 3.** The electronic band structures of **(a)** R-C2-graphyne, **(b)** R-C4-graphyene, **(c)** R-C2-BNyne, and **(d)** R-C4-BNyne.

**Figure 4.** **(a)** The electron transmission spectra of GRP-$m$GRPY2-GRP and **(b)** GRP-$m$GNRY4-GRP devices ($m$ = 1−3).

**Figure 5.** **(a)−(c)** $I$−$V$ curves of GRP-1GRPY2-GRP, GRP-2GRPY2-GRP, and GRP-3GRPY2GRP devices, respectively ($V_{bias}$ = 50−150 mV). **(d)** $I$−$V$ curves (linear scale) with ON/OFF ratios for GRP-$m$GRPY2-GRP devices ($m$ = 1−3, $V_{bias}$ = 50 mV).

**Figure 6.** **(a)−(c)** $I$−$V$ curves of GRP-1GRPY4-GRP, GRP-2GRPY4-GRP, and GRP-3GRPY4GRP devices, respectively ($V_{bias}$ = 50−150 mV). **(d)** $I$−$V$ curves (linear scale) with ON/OFF ratios for GRP-$m$GRPY4-GRP devices ($m$ = 1−3, $V_{bias}$ = 50 mV).

**Figure 7.** The structures of **(a)** GRP-$m$BNY2-GRP and **(b)** GRP-$m$BNY4-GRP devices ($m$ = 1−3). Grey, white, bronze, blue atoms are carbon, hydrogen, boron, and nitrogen atoms, respectively. The unit size of BNyne is defined in a same manner used in $R$-graphyne based devices as shown in Figure 2.

**Figure 8.** The electron transmission spectra of (**a**) GRP-*m*BNY2-GRP and (**b**) GRP-*m*BNY4-GRP devices (*m* = 1−3). *I*−*V* curves with ON/OFF ratios of (**c**) GRP-*m*BNY2-GRP and (**d**) GRP-*m*BNY4-GRP devices (*m* = 1−3, $V_{bias}$ = 1.0 V).

**Figure 9. (a,b)** LUMO and HOMO of GRP-1GRPY2-GRP device, **(c,d)** LUMO and HOMO of GRP-3GRPY2-GRP device, and **(e,f)** LUMO and HOMO of GRP-3GRPY4-GRP device with an isovalue of 0.05 Å $^{-3/2}$.

**Figure 10. (a,b)** LUMO and HOMO, **(c,d)** LUMO-1 and HOMO-1, and **(e,f)** LUMO-2 and HOMO-2 of GRP-3BNY2-GRP device with an isovalue of 0.01 Å $^{-3/2}$.

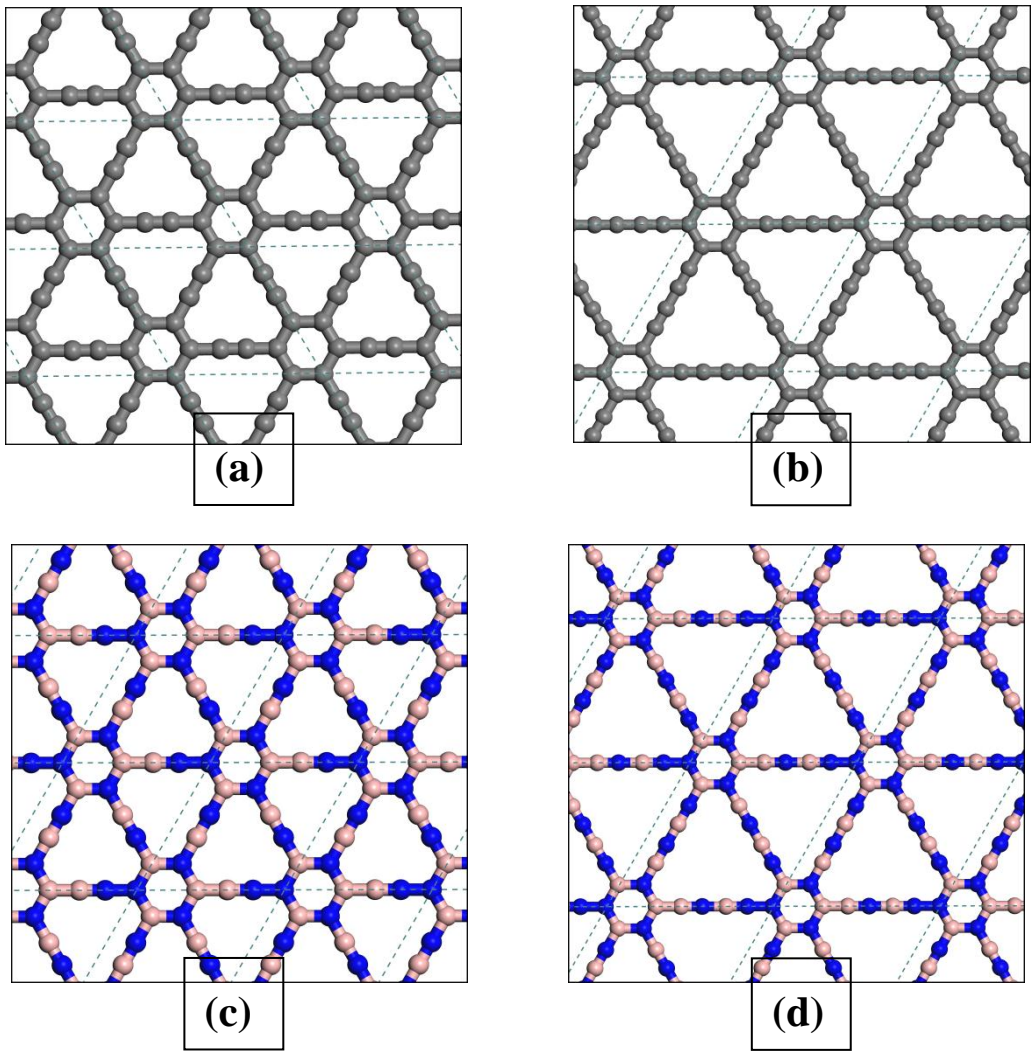

**Figure 1.** The structures of **(a)** R-C2-graphyne, **(b)** R-C4-graphyene, **(c)** R-C2-BNyne, and **(d)** R-C4-BNyne. Gray, bronze, and blue atoms are carbon, boron, and nitrogen atoms, respectively. The region surrounded by green dotted lines indicates a primitive cell of each system.

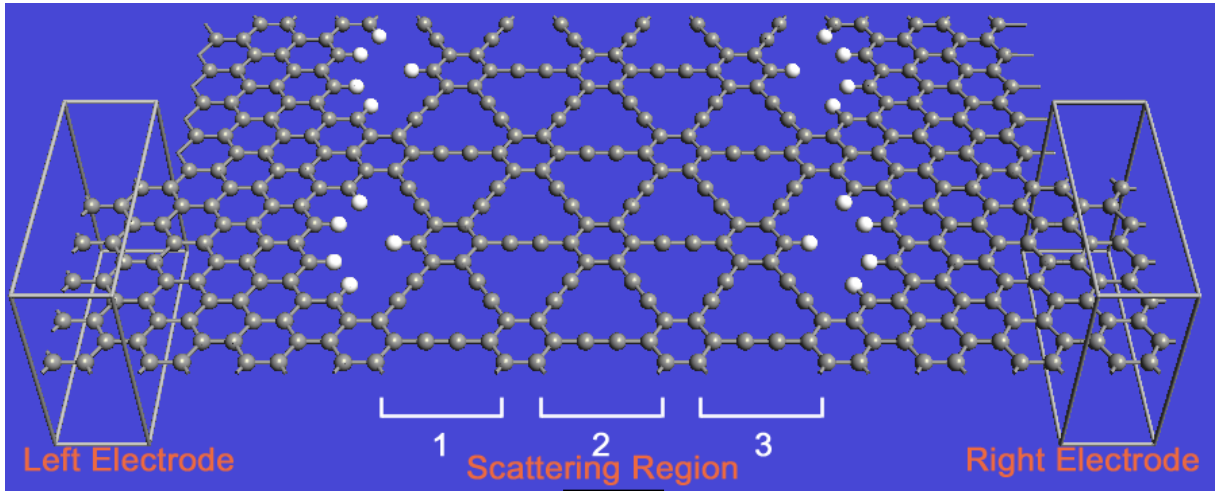

**(a)**

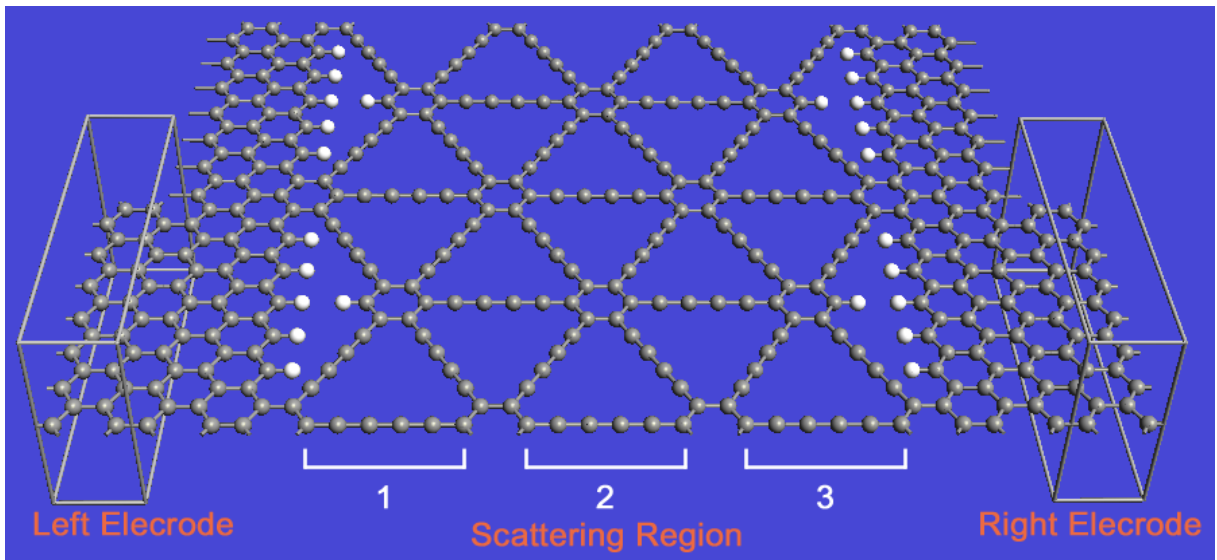

**(b)**

**Figure 2.** The structures of **(a)** GRP-*m*GRPY2-GRP and **(b)** GRP-*m*GRPY4-GRP devices (*m* = 1−3). Grey and white atoms are carbon and hydrogen atoms, respectively. The unit size of *R*-graphyne is denoted by white clamp notation and its number (*m*) ranges over 1−3.

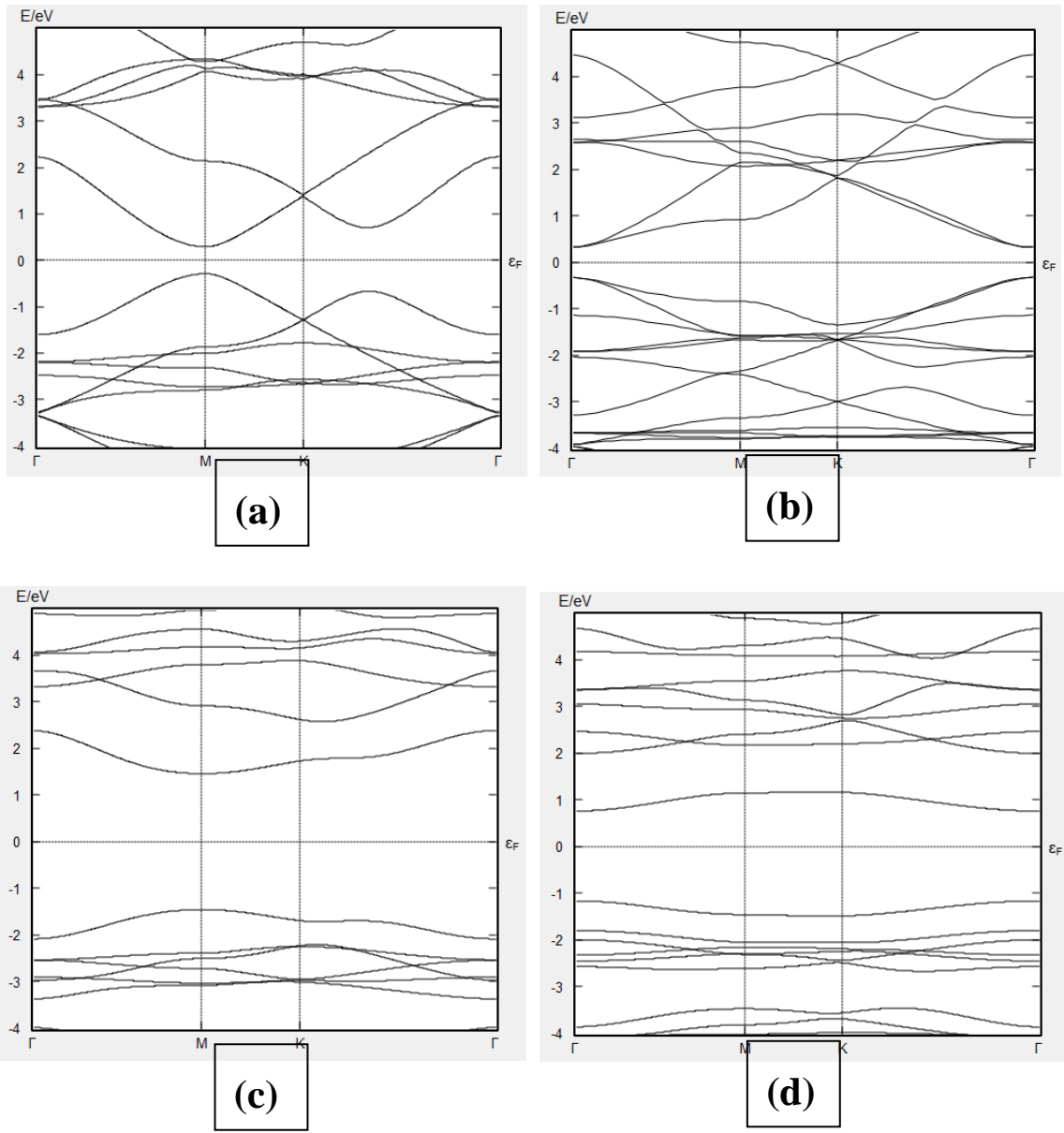

**Figure 3.** The electronic band structures of **(a)** R-C2-graphyne, **(b)** R-C4-graphyene, **(c)** R-C2-BNyne, and **(d)** R-C4-BNyne.

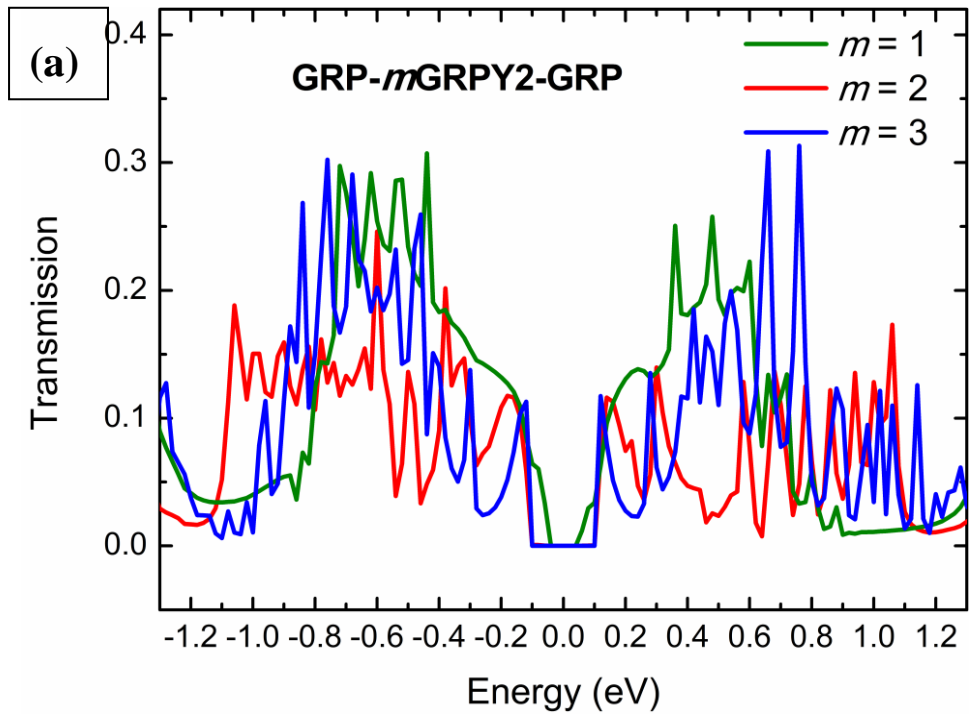

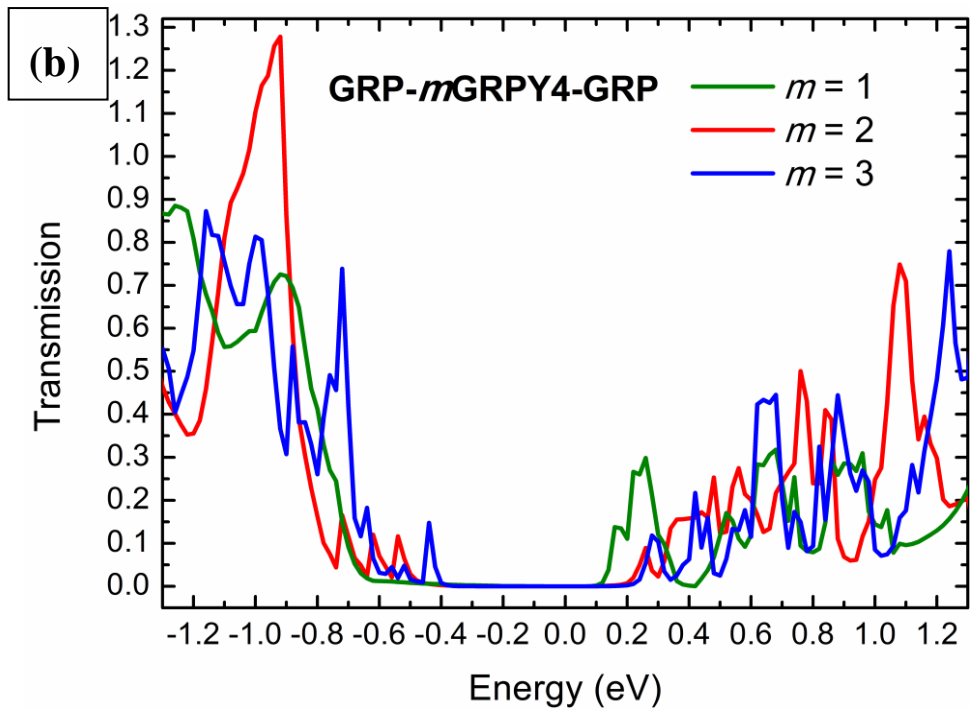

**Figure 4. (a)** The electron transmission spectra of GRP-*m*GRPY2-GRP and **(b)** GRP-*m*GNRY4-GRP devices (*m* = 1−3).

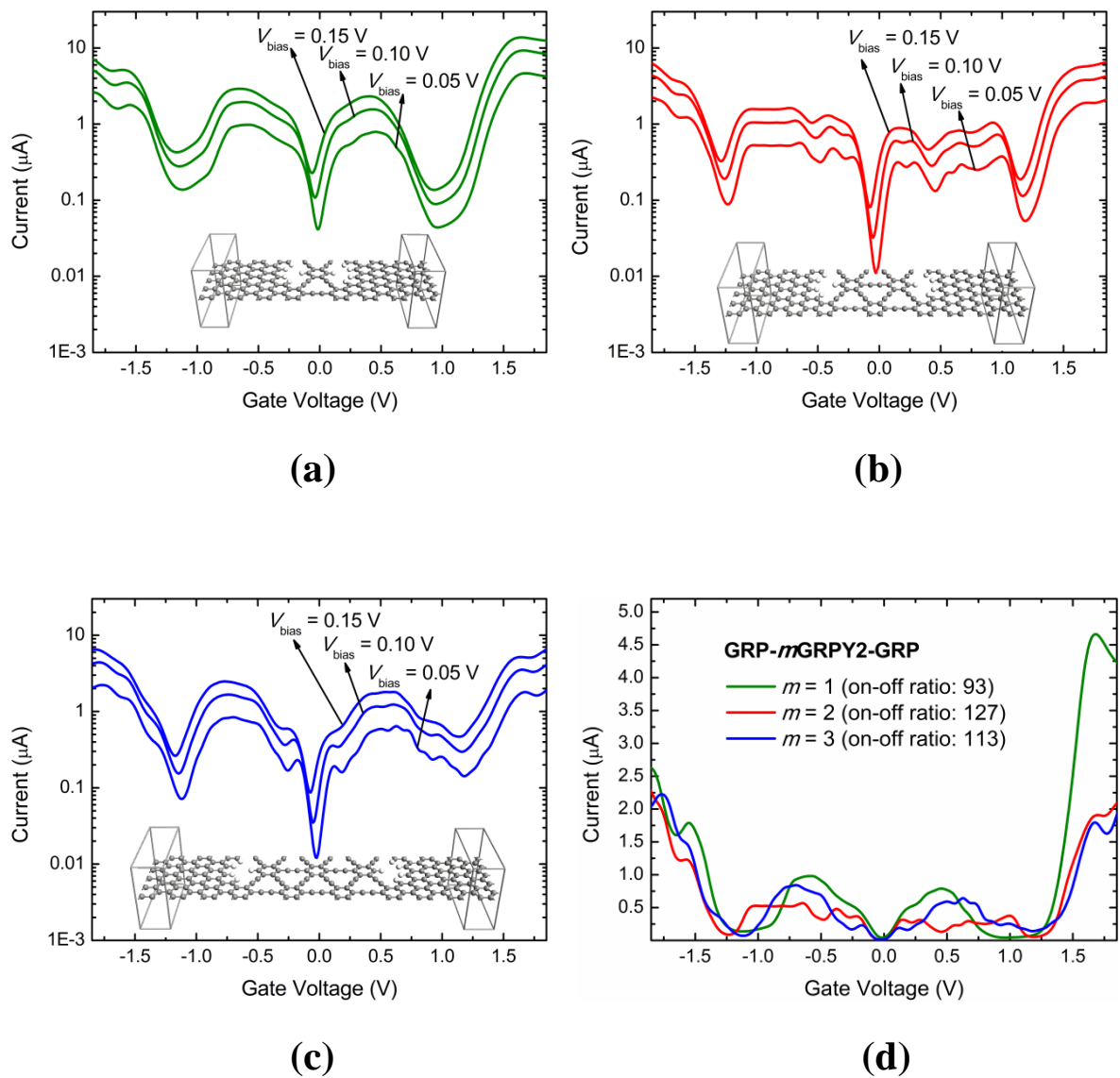

**Figure 5.** (a)−(c) *I*−*V* curves of GRP-1GRPY2-GRP, GRP-2GRPY2-GRP, and GRP-3GRPY2GRP devices, respectively. ($V_{bias}$ = 50−150 mV). (d) *I*−*V* curves (linear scale) with ON/OFF ratios for GRP-*m*GRPY2-GRP devices (*m* = 1−3, $V_{bias}$ = 50 mV).

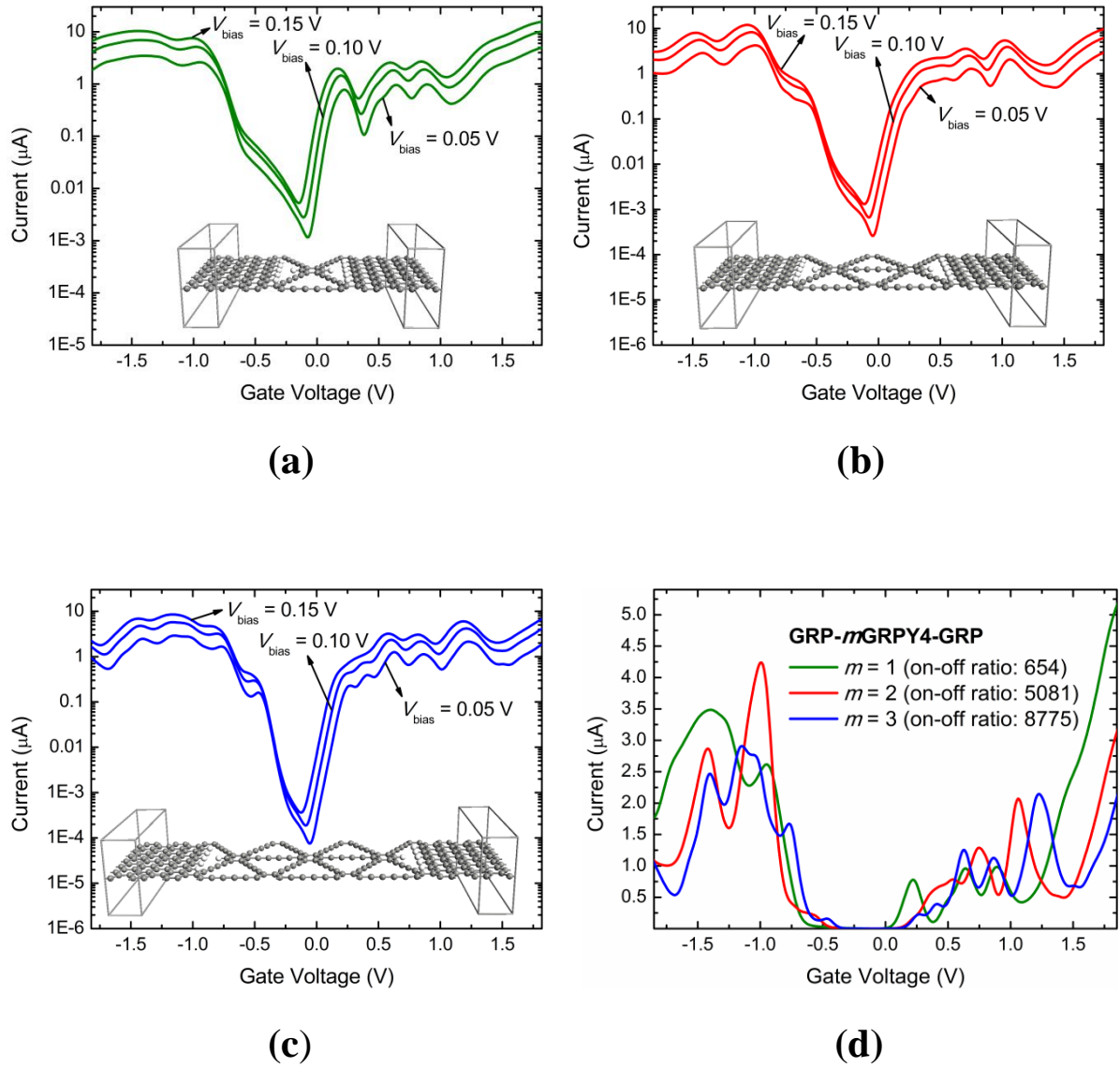

**Figure 6.** (a)−(c) *I*−*V* curves of GRP-1GRPY4-GRP, GRP-2GRPY4-GRP, and GRP-3GRPY4GRP devices, respectively ($V_{\text{bias}}$ = 50−150 mV). (d) *I*−*V* curves (linear scale) with ON/OFF ratios for GRP-*m*GRPY4-GRP devices (*m* = 1−3, $V_{\text{bias}}$ = 50 mV).

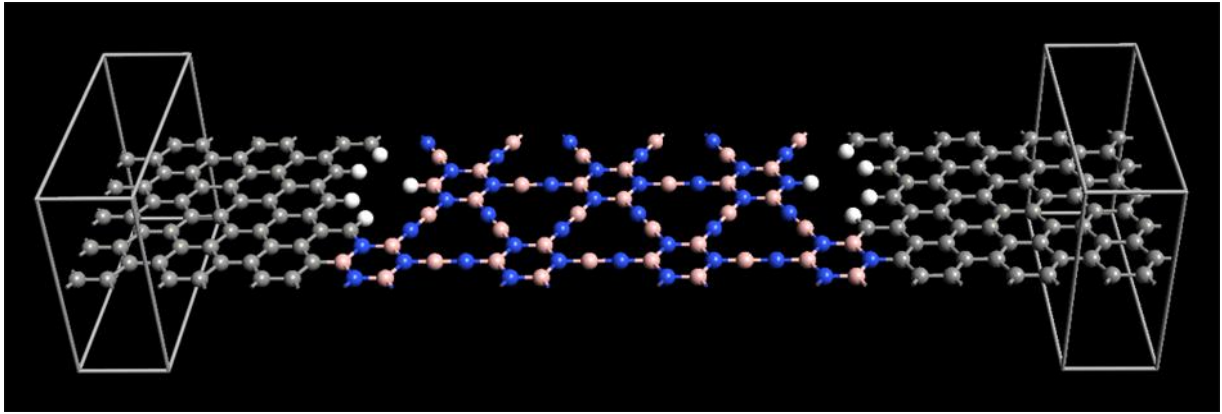

**(a)**

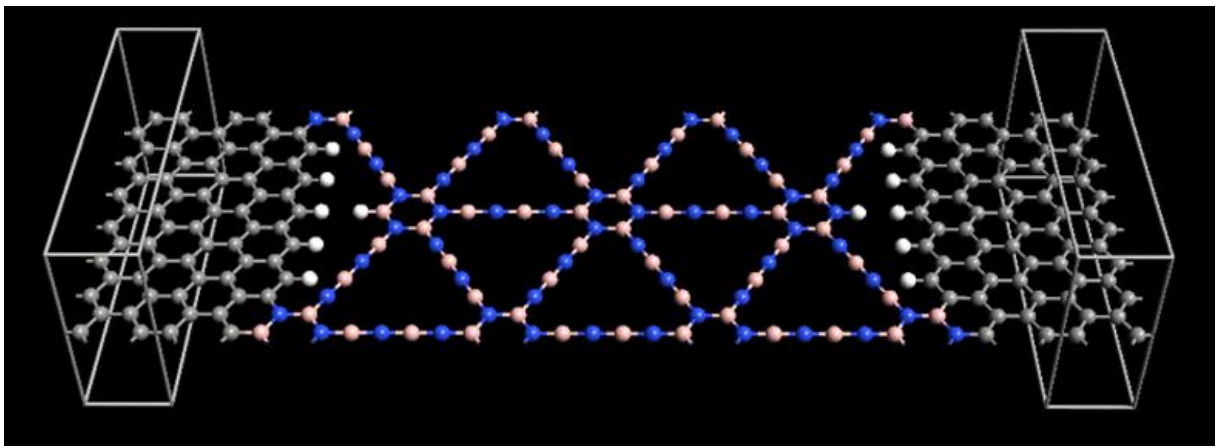

**(b)**

**Figure 7.** The structures of **(a)** GRP-*m*BNY2-GRP and **(b)** GRP-*m*BNY4-GRP devices (*m* = 1−3). Grey, white, bronze, blue atoms are carbon, hydrogen, boron, and nitrogen atoms, respectively. The unit size of BNyne is defined in a same manner used in *R*-graphyne based devices as shown in Figure 2.

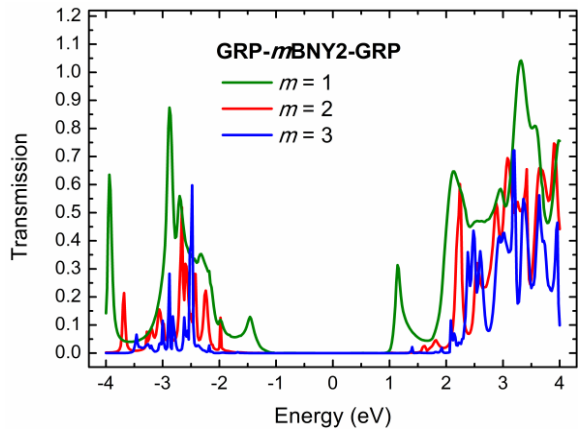

**(a)**

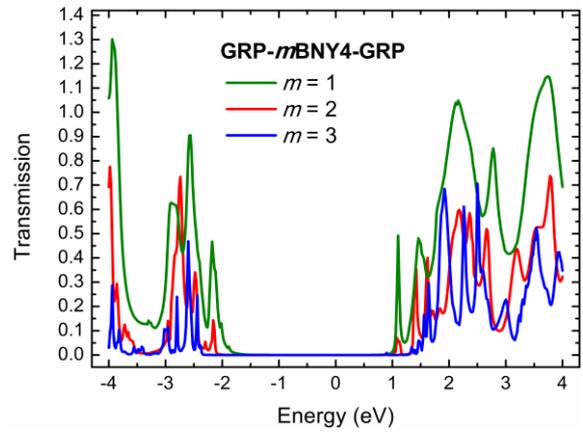

**(b)**

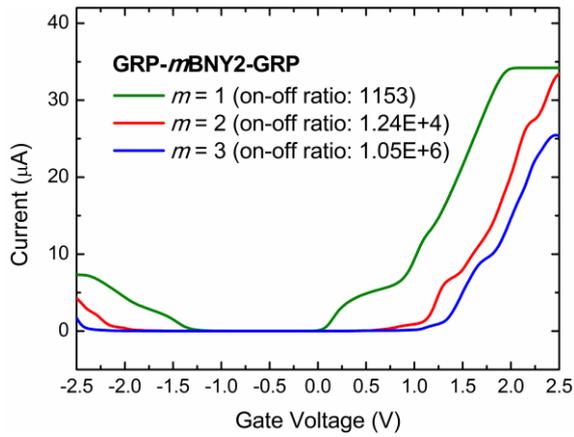

**(c)**

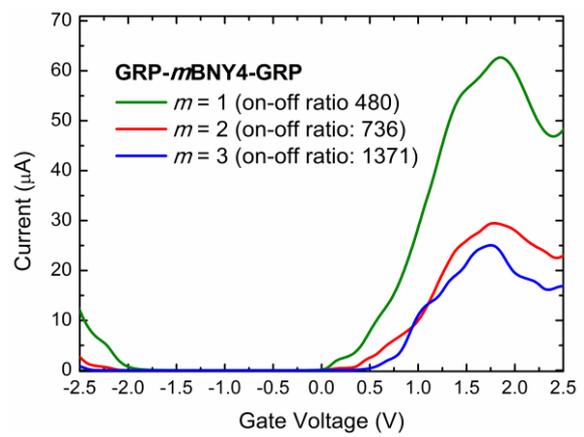

**(d)**

**Figure 8.** The electron transmission spectra of (**a**) GRP-*m*BNY2-GRP and (**b**) GRP-*m*BNY4-GRP devices ($m = 1-3$). $I-V$ curves with ON/OFF ratios of (**c**) GRP-*m*BNY2-GRP and (**d**) GRP-*m*BNY4-GRP devices ($m = 1-3$, $V_{bias} = 1.0$ V).

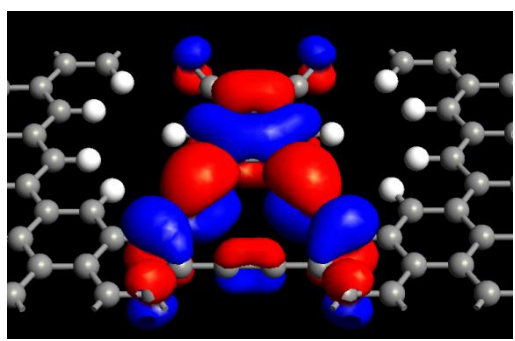
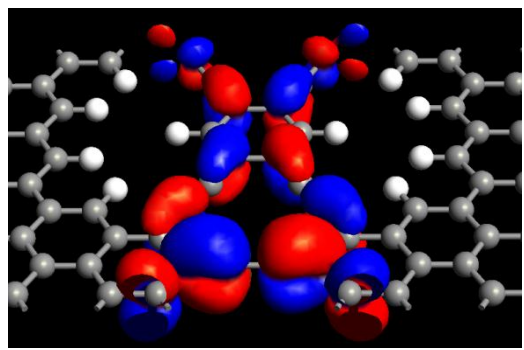

**(a)**                                      **(b)**

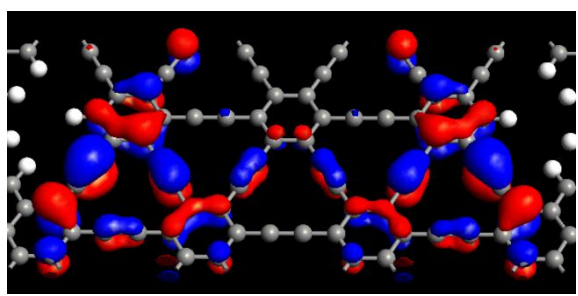
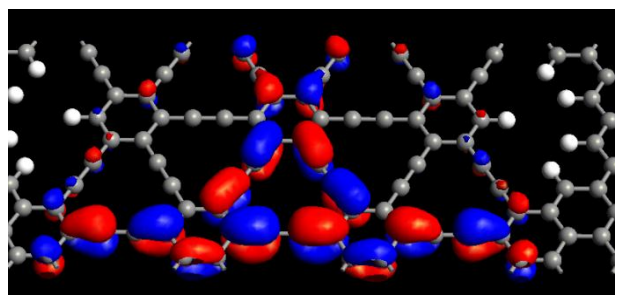

**(c)**                                      **(d)**

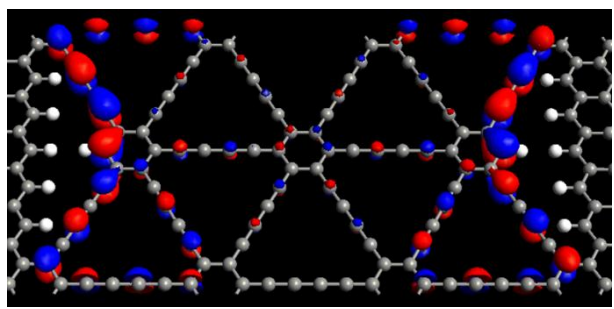
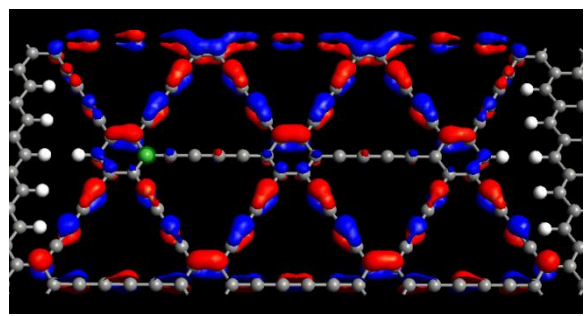

**(e)**                                      **(f)**

**Figure 9. (a,b)** LUMO and HOMO of GRP-1GRPY2-GRP device, **(c,d)** LUMO and HOMO of GRP-3GRPY2-GRP device, and **(e,f)** LUMO and HOMO of GRP-3GRPY4-GRP device with an isovalue of 0.05 Å$^{-3/2}$.

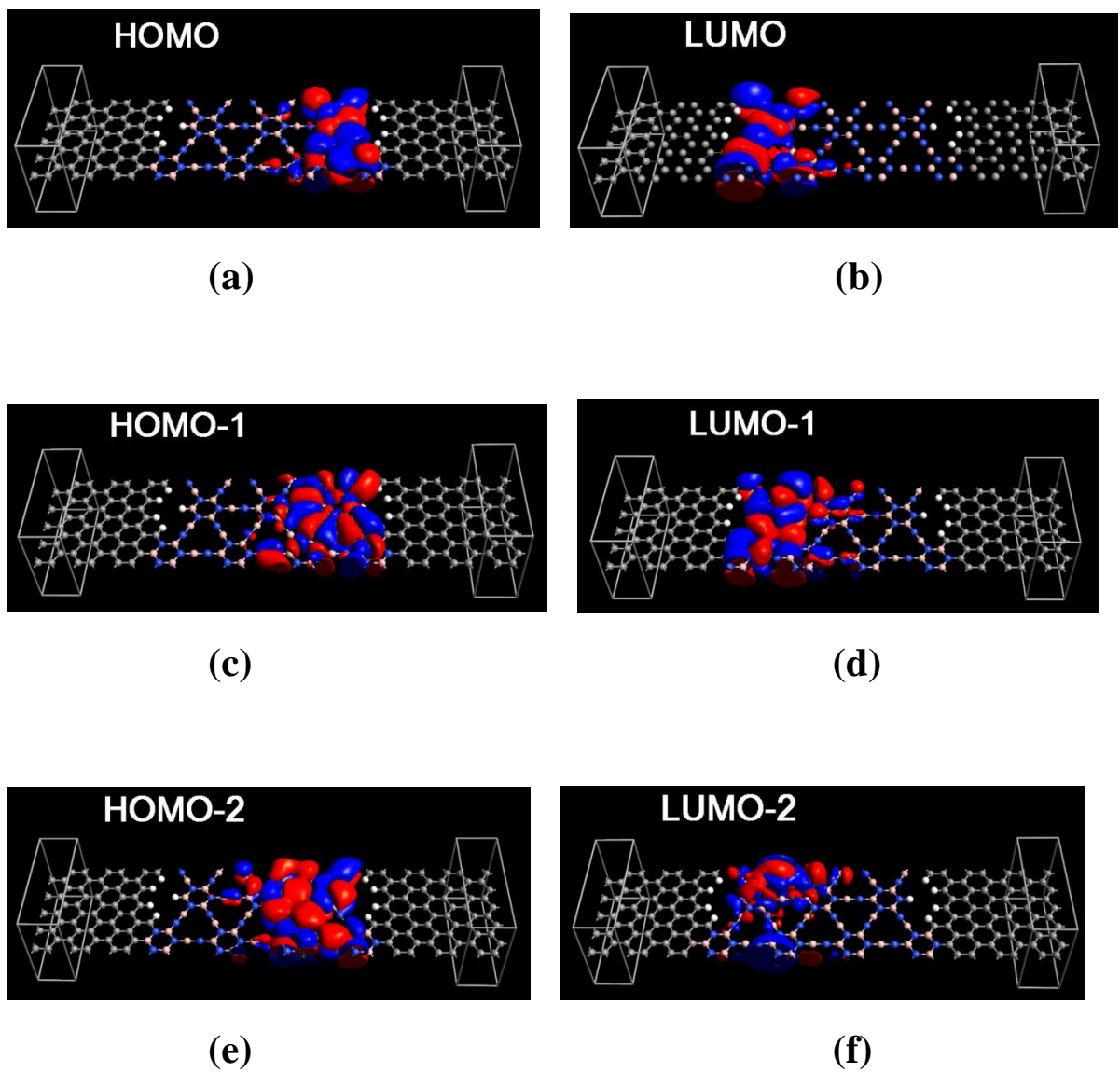

**Figure 10. (a,b)** LUMO and HOMO, **(c,d)** LUMO-1 and HOMO-1, and **(e,f)** LUMO-2 and

HOMO-2 of GRP-3BNY2-GRP device with an isovalue of 0.01 Å $^{-3/2}$.

**Table of Content**

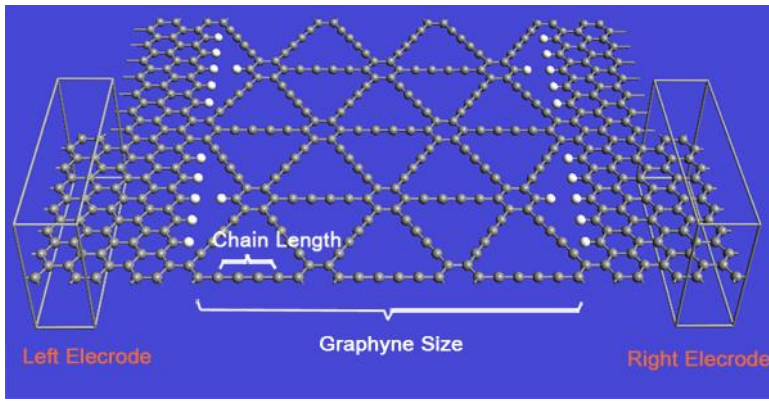 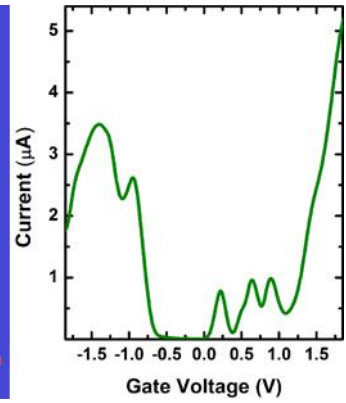